\documentclass[journal,twocolumn,letterpaper]{IEEEJERM}
  
\usepackage{times,amsmath,epsfig}
\usepackage{fancyhdr}
\usepackage{amsmath}
\usepackage{amsfonts}
\usepackage{amssymb}
\usepackage[latin1]{inputenc}
\usepackage{array}
\usepackage{graphicx}
\usepackage{url}
\usepackage{bm}
\usepackage{breqn}
\usepackage{xcolor}
\usepackage{soul}
\usepackage{gensymb}
\usepackage{subfig}
\usepackage{graphicx}
\usepackage{cite}
\begin{document}

%
% paper title
% Titles are generally capitalized except for words such as a, an, and, as,
% at, but, by, for, in, nor, of, on, or, the, to and up, which are usually
% not capitalized unless they are the first or last word of the title.
% Linebreaks \\ can be used within to get better formatting as desired.
% Do not put math or special symbols in the title.
\title{Simultaneous Monitoring of Multiple People's Vital Sign Leveraging a Single Phased-MIMO Radar}

%
% author names and IEEE memberships
% note positions of commas and nonbreaking spaces ( ~ ) LaTeX will not break
% a structure at a ~ so this keeps an author's name from being broken across
% two lines.
\author{Zhaoyi~Xu, Cong~Shi, Tianfang~Zhang, Shuping~Li, Yichao~Yuan,~\IEEEmembership{Student~Member,~IEEE,} Chung-Tse Michael~Wu,~\IEEEmembership{Senior~Member,~IEEE,} Yingying~Chen,~\IEEEmembership{Fellow,~IEEE,} Athina~Petropulu,~\IEEEmembership{Fellow,~IEEE}%
% \thanks{This work was supported by NSF under grant ECCS-2033433}
}% 

% The paper headers
\markboth{IEEE Journal of Electromagnetics, RF and Microwaves in Medicine and Biology}
{Z. Peng \MakeLowercase{\textit{et al.}}: Demo of IEEE Journal of Electromagnetics, RF and Microwaves in Medicine and Biology (JERM)}

\twocolumn[
\begin{@twocolumnfalse}
  
% make the title area
\maketitle

% As a general rule, do not put math, special symbols or citations
% in the abstract or keywords.
\begin{abstract}
Vital sign monitoring plays a critical role in tracking the physiological state of people and enabling various health-related applications (e.g., recommending a change of lifestyle, examining the risk of diseases). Traditional approaches rely on hospitalization or body-attached instruments, which are costly and intrusive. Therefore, researchers have been exploring contact-less vital sign monitoring with radio frequency signals in recent years. Early studies with continuous wave radars/WiFi devices work on detecting vital signs of a single individual, but it still remains challenging to simultaneously monitor vital signs of multiple subjects, especially those who locate in proximity. In this paper, we design and implement a time-division multiplexing (TDM) phased-MIMO radar sensing scheme for high-precision vital sign monitoring of multiple people. Our phased-MIMO radar can steer the mmWave beam towards different directions with a micro-second delay, which enables capturing the vital signs of multiple individuals at the same radial distance to the radar. Furthermore, we develop a TDM-MIMO technique to fully utilize all transmitting antenna (TX)-receiving antenna (RX) pairs, thereby significantly boosting the signal-to-noise ratio. Based on the designed TDM phased-MIMO radar, we develop a system to automatically localize multiple human subjects and estimate their vital signs. Extensive evaluations show that under two-subject scenarios, our system can achieve an error of less than 1 beat per minute (BPM) and 3 BPM for breathing rate (BR) and heartbeat rate (HR) estimations, respectively, at a subject-to-radar distance of $1.6~m$. The minimal subject-to-subject angle separation is $40\degree$, corresponding to a close distance of $0.5~m$ between two subjects, which outperforms the state-of-the-art. 

\end{abstract}

% Note that keywords are not normally used for peerreview papers.

\end{@twocolumnfalse}]

% Put footnotes here
{
  \renewcommand{\thefootnote}{}%
  \footnotetext[1]{This work was supported by NSF under grant ECCS-2033433}
%   \footnotetext[2]{Authors' affiliation 2}
}
 
% For peer review papers, you can put extra information on the cover
% page as needed:
% \ifCLASSOPTIONpeerreview
% \begin{center} \bfseries EDICS Category: 3-BBND \end{center}
% \fi
%
% For peerreview papers, this IEEEtran command inserts a page break and
% creates the second title. It will be ignored for other modes. 
\IEEEpeerreviewmaketitle

% \section{Introduction}
% The very first letter is a 2 line initial drop letter followed
% by the rest of the first word in caps.
% 
% form to use if the first word consists of a single letter:
% \IEEEPARstart{A}{demo} file is ....
% 
% form to use if you need the single drop letter followed by
% normal text (unknown if ever used by the IEEE):
% \IEEEPARstart{A}{}demo file is ....
% 
% Some journals put the first two words in caps:
% \IEEEPARstart{T}{his demo} file is ....
% 
% Here we have the typical use of a "T" for an initial drop letter
% and "HIS" in caps to complete the first word.

\section{Introduction}

\IEEEPARstart
%{N}{owadays}, people are spending more time doing sedentary activities in indoor venues (e.g., working in corporate offices, watching TV at home), which leads to many health issues due to their inactivity. 
{T}{racking} of the physiological states of people can enable change of lifestyle recommendations from indoor sedentary activities and examine the risk of diseases.
% viral infections, such as COVID-19. 
% It is important to track the physiological state of people to suggest a change of lifestyle, examine the risk of diseases, and even detect viral infections, such as COVID-19.
Vital signs, including breathing rate (BR) and heartbeat rate (HR), provide crucial insights into the physiological state of the individual. Traditional ways to monitor vital signs usually require hospitalization and involve body-attached instruments (e.g., PPG and ECG sensors), which are intrusive, costly, and require the cooperation of the person being monitored. To overcome these problems, research studies have been exploring contact-less vital sign monitoring via radio frequency (RF) signals~\cite{wang2020review, yuan2020high, rahman2015signal, gu2013hybrid, adib2015smart}. Early studies have leveraged either continuous-wave radar or a WiFi device to transmit RF signals towards a target individual, and then infer the chest movements caused by vital signs through analyzing the echo signals~\cite{liu2015tracking, mercuri2017frequency}. 
% \textcolor{red}{However, these methods usually rely on RF signals operating at fixed frequencies, which limit their capabilities on disentangling echo signals from multiple individuals, especially those who locate in proximity.}
However, these methods rely on RF signals operating at fixed frequencies, and thus have limited ability to disentangle echo signals from multiple individuals, especially when those individuals are closely located with respect to each other. 
Such limitation precludes their use in practical scenarios that involve the health monitoring of multiple people. For example, in a classroom or a clinic, under the COVID-19 crisis, tracking the vital signs and health status of multiple people simultaneously in a contactless fashion would be very desirable. 
% , such as people working side by side in an office, or watching TV together. 
% It is highly desired to have a contact-less system that can track the vital signs and health status of multiple people simultaneously.    \textcolor{red}{why? I think a better motivational example is needed. **** Maybe tracking multiple individuals would be important in overflowing emergency rooms, or ICU's, as we saw during the covid crisis.}

% signs, including breathing rate and heartbeat rate, are important metrics in asscessing the physical health of a person. 
% These vital signs provide crucial insights into the psychological state of the individual, such as sleep stages and stress levels, which have major impacts on the human healthcare system. Vital signs monitoring also enables timely diagnosis and treatment of various respiration and cardiovascular diseases, such as sleep apnea, asthma, lung disorder, and heart failure. 
 
% Such a limitation precludes many real-world scenarios that involve multiple target individuals, such as couples that share the same bed and parents who sleep with their children. 
% Furthermore, the low-frequency RF signals (usually below 2.4GHz) used in these approach only achieve centimeter-level tracking, making it difficult to precisely capture minor heartbeats. 
% A system that is capable of simultaneously monitoring the vital signs of multiple people at high precision is highly desired.   

Powered by recent advancements in mmWave sensing, research studies have been exploring mmWave radars for vital sign monitoring~\cite{ahmad2018vital, alizadeh2019remote, sakamoto2019noncontact, islam2020non, vodai2021enhancement, Lv2021Noncontact,8777845, Turppa2020vitalsign}. By utilizing frequency-modulated continuous-waveform (FMCW) techniques, a mmWave radar can detect multiple people at different radial distances to the radar device, and further derive vital signs information of each individual. Compared to traditional approaches relying on low-frequency RF signals, mmWave signals have much shorter wavelength (i.e., millimeter-level wavelength) and thus can enable more fine-grained vital sign monitoring. 
%To provide enhanced sensing performance with mmWave, 

Two main categories of techniques were explored, namely, analog beamforming based on phased antenna array and multiple-input and multiple-output (MIMO) approaches.  
% The beamforming approach of ~\cite{ahmad2018vital, islam2020non} can improve the directivity of an antenna array by iteratively \textcolor{red}{successively??} steering the mmWave beam towards different angles. 
The beamforming approach of ~\cite{ahmad2018vital, islam2020non} uses a phased array to successively steer the mmWave beam towards different directions by changing the antenna weights in each slot.  
% **** Athina: is this ok??****
This approach enables the detection of people separated in the angle domain, while the allowable minimum angle separation (resolution) is limited by the phased array's aperture. 
% while the allowable minimum angle separation (resolution) decreases as the phased array aperture increases. %
By processing the echoed signals of each beam separately, the vital signs of the person in each direction can be estimated.
Specifically, Islam~\textit{et al.}~\cite{islam2020non} deploy analog beamforming  on a single-channel FMCW radar to measure vital signals of two subjects with a minimal 40 degree separation angle. 
% \textbf{ychen: Need to say what accuracy this work achieved. Otherwise, this sentence indicated that minimal 40 degree angle separation is done already. There is no novelty in our approach.}  
However, this work only reported preliminary BR and HR estimation performance of a single subject (i.e., around $93\%$ HR estimation and $96\%$ BR estimation accuracy), while the multi-subject vital sign estimation precision and interference among subjects were not studied.
%Another way to improve the performance of vital sign monitoring is
MIMO approaches for 
 vital sign monitoring 
~\cite{alizadeh2019remote, vodai2021enhancement} transmit mmWave signals using multiple transmitting antenna (TX) - receiving antenna (RX) pairs, in a time-division-multiplexing (TDM) fashion, to realize MIMO and improve sensing precision.
% leverage multiple orthogonal waveforms {so they can transmit multiple waveforms?}, transmitted in a  TDM fashion, to improve  sensing precision. 
For example, Dai~\textit{et al.}~\cite{vodai2021enhancement} exploit a MIMO radar with 12 TXs and 16 RXs.  In each time slot, only one of the available TX transmits. 
% {then why do they use orthogonal waveforms?}, 
Over 12 slots, measurements corresponding to 192 TX-RX antenna pairs/channels can be collected and used for high-precision vital sign monitoring with low estimation error. 
% \textbf{Again, you need to say what estimation accuracy or error they can achieve. And state clearly is this for multiple people or individual. Otherwise, reviewers feel very low estimation error has already been achieved.} 
However, the designed method was only evaluated with multiple people at different radial distances to the radar, while the more challenging scenario of multiple people at the same radial distance but different angles, was not studied. All these existing solutions consider either analog beamforming or MIMO techniques, but not both, thus missing the opportunity to  fully explore the potential of radar-based mmWave sensing.
% ****Athina: is this what is happening???*** 
% \textcolor{red}{Does this use 12 different RF chains?, i.e., 12 different waveforms? or they do it in a TDM fashion?}{Xu: No, they still do it in a TDM fashion. I checked TI's user guide on this board, TI only allows this board to do TDMA MIMO because this board still only has 1 RF chain.} 
% 

% \textcolor{red}{******this is not true. If they use a virtual array they can achieve high resolution even without beamforming. Beamforming allows us to direct the transmit power to a certain direction - it does not affect the resolution.****}{Xu: But they did not estimate angle at all. They only separate targets by range. This gives us chance to finish this paper.} 

In this paper, we propose a novel approach for high-precision  vital sign monitoring of multiple people.  In particular, we integrate MIMO and phased array by using a single-chip automotive mmWave transceiver and develop a  phased-MIMO radar sensing scheme. %
The radar   transmits   orthogonal signals,  each  feeding  a  phase  array  structure.  Orthogonality allows the transmitted signals to be separated at the receiver\cite{Sun2020MIMO}. The contributions of the multiple waveforms offer independent views of the targets, which can be exploited to improve target  estimation.  In  our  work,  orthogonality  is  achieved  by time-division multiplexing (TDM) transmission  of  the  same  waveform  after it  has been weighted by   different weights in each slot. The received signals over multiple time slots can be combined to obtained high precision target estimation.
It is those different transmitted signals that makes our work different than that of \cite{vodai2021enhancement}, resulting in more transmitted-received signal pairs for more accurate target estimation.
% which can be exploited for more accurate target estimation.

Our phased-MIMO radar can steer the mmWave beam towards different directions with a micro-second delay, which enables simultaneously monitoring the vital signs of multiple individuals even when they are
% {almost simultaneously.} \sout{
at the same radial distance to the radar.
% \textcolor{red}{I do not understand why this allows monitoring of subjects at the same radial distance  and not at any distance. (Athina)} }
% \sout{
% Furthermore, we develop a time-division multiplexing (TDM)-MIMO technique to boost the signal-to-noise ratio, by fully utilizing all TXs and RXs of the radar. The phased-MIMO radar gives us more degrees of freedom (than a phased array) to design the beampattern. It also allows us to construct a virtual array for enhanced angle resolution.}
The phased-MIMO radar also allows one to construct a virtual array for enhanced angle resolution; this was verified in our preliminary experiments detecting stationary objects. 
% {but this seems to not have been tested  (Athina)}
By combining high angle resolution and high precision target estimation, the proposed system enables the detection of  multiple people located close to each other, and also the precise estimation of the {BR} and {HR} of each individual. Our approach thus provides a promising
solution to track the health status of multiple people in many indoor venues (e.g., classrooms, offices, and crowded hospital rooms).

Based on the proposed radar, we develop a system to automatically localize multiple human subjects and estimate their vital signs. Our system first leverages FMCW techniques to obtain the angle of each individual subject to the radar. 
% With the large aperture enabled by MIMO, our system can achieve higher angle estimation precision compared with existing phased-array based approaches. 
Then, to separate the vital signs of different people, our system iteratively steers the  beam towards each individual subject and obtains {a set of mmWave signals}. 
% \textcolor{red}{Athina: ***** the large number of degree of freedom of MIMO allow us to formulate a beam with a lobe pointing to each source signal of interest simultaneously. However, by doing it in TDM, in each each time slot, we do analog beamforming, having the same issues as ~\cite{ahmad2018vital, islam2020non}  ***8} 
Each set of mmWave signals thus contain vital signals of only one single subject. For each subject, our system computes the phase of the mmWave signals, which encodes both the breathing and heartbeats of the subject. Two band-pass filters, which use normal human breathing and heartbeat frequency ranges as cut-off frequencies, are employed to separate the two types of vital signs. Our system then detects the BR and HR in the frequency domain by locating the frequency peaks. 
% To detect heart rate, our algorithm first applies a bandpass filter to eliminate irrelevant frequency components, and then estimates the heart rate in the frequency domain by locating the frequency peak in the normal heart rate range.
% Our system then computes the range FFT to obtain the radial distance between the radar and each subject. 
We implement the designed TDM phased-MIMO radar on an off-the-shelf mmWave device and conduct extensive experiments under various settings (e.g., different distances and angles between the radar and subjects). The results show that our system can provide high accuracy BR and HR estimation under various experimental settings.

% In this paper, we explore integrating both MIMO and phased array techniques into a single-chip automotive mmWave device and develop a TDM Phased-MIMO radar. Our mmWave radar can steer the mmWave beam towards different individuals with a micro-second delay, which facilitates extracting the vital signs of multiple individuals simultaneously. 
% \textbf{Comparing our performance with state-of-the-art method here XX}

% \textbf{Add paragraph 4: talk about the techniques} 

% - how to separate the vital sign of two people
 
% - how to extract measurements containing heartbeat and breathing of each individual
 
% - how to separate the heartbeat rate and breathing rate

% - the study we conducted, distance study, angle study for evaluation

% \textbf{some post-processing techniques we developed XXXX}
     
\vspace{-1mm}
\section{vital sign detection with phased-MIMO radar}

% \vspace{-2mm}
\subsection{TDM Phased-MIMO}
{Phased-MIMO radar is the combination of MIMO radar with phased array, where the radar transmits orthogonal signals, each feeding a phase array structure.
 Orthogonality allows the transmitted signals to be separated at the receiver. The contributions of the multiple waveforms offer independent views of the targets, which can be  exploited to improve target estimation.
 In our work, orthogonality  is achieved by TDM transmission of the same waveform that is weighted by different weights in each slot.}
% achieved by either frequency division multiplexing(FDM), time division multiplexing(TDM), code division multiplexing(CDM) and other methods.

First, let us consider a transmitting array where we have $N$ {TXs} spaced by $d_t$, and a receiving array with $M$ {RXs} spaced by $d_r$.  {The transmitting array is partitioned into $P$ overlapped subarrays where antennas can be shared between subarrays. In each time slot, the subarrays take turns to 
 transmit the same waveform $x(t)$ but using different subarray antenna weights. The weights are chosen so that all transmitted signals add up coherently in a specific direction, or equivalently,  form a beam in a specific direction. By changing the weights in each slot we effectively create different channels that will provide diversity and improve target estimation.}

Considering the $p$-th subarray always contains the $p$-th {TX}, which is used as reference to form the beam, the corresponding beamforming weights are
\begin{equation} 
\begin{aligned}
    \mathbf{w}_p = [s_0e^{j2\pi p\alpha(\theta)},\dots,1,
    % s_{k+1}e^{-j2\pi \alpha(theta)},
    \dots,s_{N-1}e^{-j2\pi (N-p-1)\alpha(\theta)}]^T
\end{aligned}
\end{equation}
where $\alpha(\theta) = d_t \frac{\sin(\theta)}{\lambda}$, $\lambda$ is the wavelength of signal and $s_l$ is $1$ if the $l$-th antenna belongs to the $p$-th subarray, otherwise is $0$.
The signal transmitted by the $p$-th array towards direction $\theta$ can be written as
\begin{equation} 
\begin{aligned}
    z_p(t) = \mathbf{a}^H_t \mathbf{w}_p x(t) = e^{j2\pi p\alpha(\theta)}x(t)\sum_{n=0}^{N-1}s_n, 
\end{aligned}
\end{equation}
where $\mathbf{a}_t = [1, e^{-j2\pi \alpha(\theta)}, \dots, e^{-j2\pi(N-1)\alpha(\theta)}]^T$ is the transmit steering vector.
One can see that the transmitted signals from $P$ subarrays are the same as those from a TDM-MIMO radar with the same array, except that each signal is amplified by the number of antennas in the corresponding subarray.
% Thus in phased-MIMO, we still enjoy the benefit of multiple channels like we do in a MIMO radar.

\vspace{-2mm}
\subsection{Vital Sign Detection with FMCW Signal}
Let us consider an FMCW radar waveform, i.e.,  
\begin{equation}
\begin{aligned}
    x(t) = A_te^{j2\pi[f_c t + \frac{B}{2T_c}t^2 + \Phi(t)]}
\end{aligned}
\end{equation}
where $A_t$ is the amplitude, $f_c$ is the chirp starting frequency, $B$ is the chirp bandwidth, $T_c$ is the chirp duration, and $\Phi(t)$ is the phase noise from transmitter. Note that the phase noise will be neglected in the following equations since it is slow varying and the propagation delay of mmWave is short.

At the radar receiver, after mixing with the conjugate of the transmitted signal, the signal transmitted by the $n$-th subarray and received by the $m$-th RX can be written as
% \begin{equation}
%     x_r(n,m,t) = \alpha A_t e^{j2\pi[f_c (t-\tau) + \frac{B}{2T_c}(t-\tau)^2 + \Phi(t-\tau) + \phi_0 + (d_n+d_m)\frac{\sin(\theta)}{\lambda}] }
% \end{equation}
% where $\alpha$ is the attenuation coefficient, $\tau = \frac{2R(t)}{c}$ is the propagation delay of an object at radial range $R(t)$, $d_n = (n-1)d_t$ and $d_m = (m-1)d_r$, respectively.
% {If the phased-MIMO is used here, the phase shift term  $e^{(d_n)\frac{\sin(\theta)}{\lambda}}$ in the following equations will be cancelled.}
%  the signal can be expressed as:
\begin{equation}
\begin{aligned}
    y(n,m,t)&= 
            % x_t^*(t)x_r(n,m,t)\nonumber\\
            % &= \alpha A_t^2 e^{-j2\pi[f_c\tau + \frac{B}{T_c}\tau t - \frac{B}{2T_c}\tau^2 + \Phi(t)-\Phi(t-\tau)] - (d_n+d_m)\frac{\sin(\theta)}{\lambda} }\nonumber\\
            % &= A_r e^{-j2\pi[\frac{2BR(t)}{cT_c}t + \frac{2f_cR(t)}{c} - \frac{2BR^2(t)}{T_cc^2} + \Delta\Phi(t)- (d_n+d_m)\frac{\sin(\theta)}{\lambda}]}\nonumber\\
            A_r e^{-j2\pi [f_bt + \textcolor{black}{\Phi_b(t,n,m)]}},
\end{aligned}
\end{equation}
where $f_b = \frac{2BR(t)}{cT_c}$ is the beat frequency, 
\begin{align*}
    % &f_b = \frac{2BR(t)}{cT_c}\\
    &\Phi_b(t,n,m) = \frac{2f_cR(t)}{c} - \frac{2BR^2(t)}{c^2T_c} - (d_m-d_n)\frac{\sin(\theta)}{\lambda},
\end{align*}
$A_r$ is the amplitude after beamforming and attenuation, $R(t)$ is the radial range of an object, which is associated and changed with the chest displacements of the target subject, $d_n = (n-1)d_t$ and $d_m = (m-1)d_r$, respectively.
Since the propagation delay is very small, the phase term
%$\Phi_b(t,n,m)$
can be approximated as $\Phi_b(t,n,m) = \frac{2f_cR(t)}{c}-(d_m-d_n)\frac{\sin(\theta)}{\lambda}$.

\begin{figure}
\centering
\includegraphics[width= 3 in]{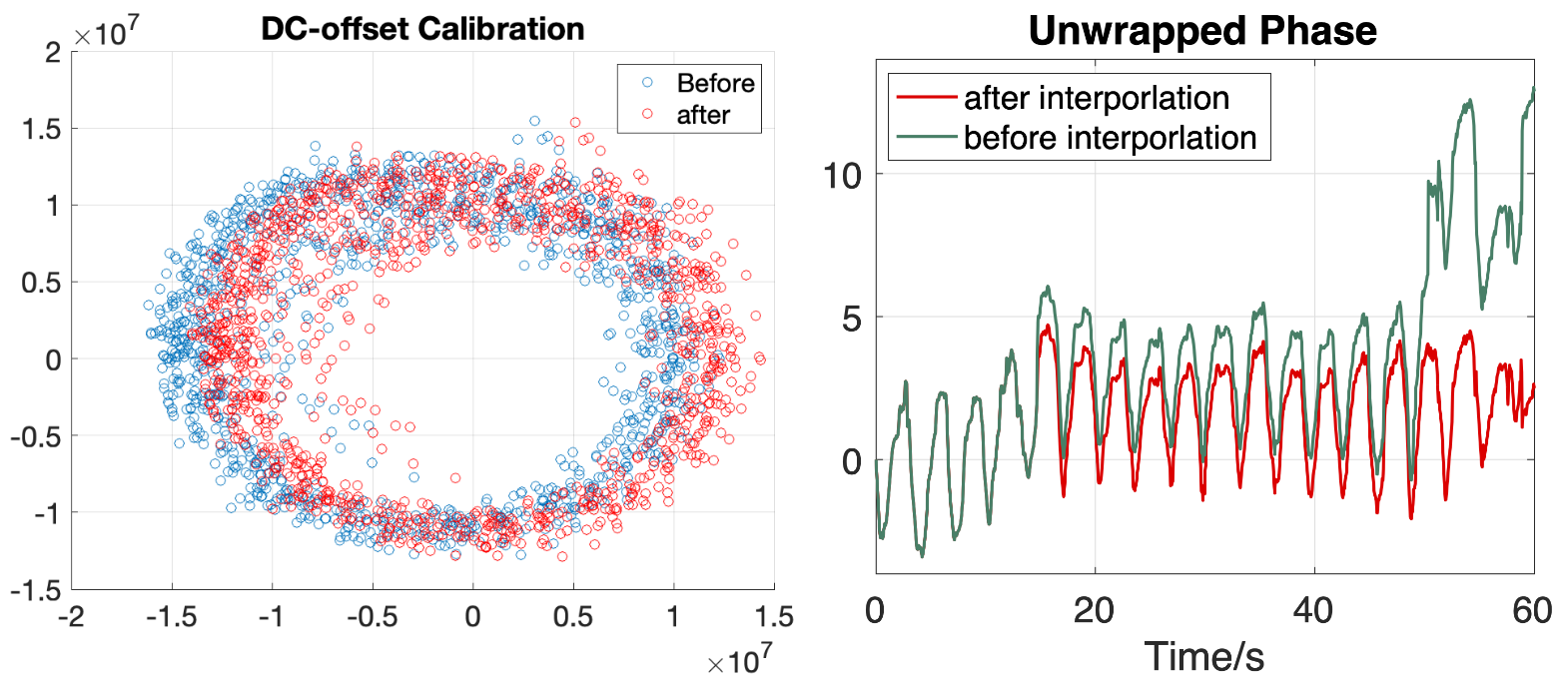}
% \begin{minipage}[t]{0.22\textwidth}
% \centering
% \includegraphics[width=1\columnwidth]{images/DC-offset.png}
% \vspace{-3mm}
% \caption{DC-offset correction using least square method}
% \vspace{-4mm}
% \label{fig:dc_offset}
% \end{minipage}
% \begin{minipage}[t]{0.22\textwidth}
% \centering
% \includegraphics[width=1\columnwidth]{images/Unwarpped_phase.eps}
\vspace{-2mm}
\caption{(a) DC-offset correction using least square method; (b) Phase drift calibration through interpolation}
\vspace{-3mm}
\label{fig:phase_unwarp}
% \end{minipage}
\end{figure}

Let the ADC sample interval be $T_f$ and the chirp interval $T_s$. The beat signal of the $k$-th ADC sample in the $l$-t chirp is
\begin{equation}
    y[n,m,k,l] = A_r e^{j2\pi[f_bkT_f + \frac{2f_c}{c}R(kT_f + lT_s) - (d_m-d_n)\frac{\sin(\theta)}{\lambda} ]}.
    \label{beat_sig}
\end{equation}
Provided that the range change due to vital sign is slow ($<2$Hz) and the sampling interval is very short, if the target stays at a nominal range $R_0$, then
% \begin{equation}
%     R(kT_f + lT_s) \approx R_0 + R_1(lT_s)
% \end{equation} 
the phase term of \eqref{beat_sig} will be
\begin{align}
   \textcolor{black}{ \Phi_b(l,n,m)} &= \frac{2}{\lambda}[R_0+R_1(lT_s)] - (d_m-d_n)\frac{\sin(\theta)}{\lambda}\nonumber\\
    &= \Phi_0(n,m)+ \frac{2R_1(lT_s)}{\lambda}.
\end{align}
and \eqref{beat_sig} can be expressed as
\begin{equation}
    y[n,m,k,l] = A_r e^{j2\pi f_bkT_f}e^{j2\pi\textcolor{black}{\Phi_b(l,n,m)}}.
    \label{beat_sig_2}
\end{equation}
On collecting  $N_s$ samples in
each chirp, the signal $y[n,m,k,l], k=1,...,N_s$ can be viewed as a complex sinusoid with frequency $f_b T_f$ and complex amplitude $e^{j2\pi\Phi_b(l,n,m)}$. Therefore, on applying an $N_s$-point Discrete Fourier Transform (DFT) on $y[n,m,k,l]$ along $k$ we
%
% there are in total $N_s$ samples collected. After applying an  $N_s$-point DFT on \eqref{beat_sig_2} within each chirp, {along dimension $k$, ***********$k$ is used for the subarray index ***** this is confusing}  we get
% \begin{align}
%     Y_{n,m,l}[h] &= \text{DFT}\{y[n,m,k,l]\} = \sum_{k=0}^{N_s-1}y[n,m,k,l]e^{-j2\pi(\frac{kh}{N_s})}\nonumber\\
%     &= A_re^{j2\pi\textcolor{black}{\Phi_b(l,n,m)}}\sum_{k=0}^{N_s-1}e^{j2\pi k(f_bT_f-\frac{h}{N_s}) }.
% \end{align}
see a peak at DFT sample $ h = N_sf_bT_f $, indicating the radial range of the target. The phase of the peak value, equals $\Phi_b(l,n,m)$, which can be used to measure the frequency of chest displacement. In this paper, we will use the real and imaginary parts of the peak corresponding to the target in the $l$-th chirp, denoted as $Y_R[l]$ and $Y_I[l]$.   
% we omit the antenna pair notation $n$ and $m$ for simplicity. 

\vspace{-2mm}
\section{Design of Vital Sign Monitoring System}

% Based on the designed TDM phased-MIMO radar, we designed a system to perform multi-people vital sign monitoring. Our system first leverage a 

\begin{figure}[t]
    \centering
    \includegraphics[width=0.44\textwidth]{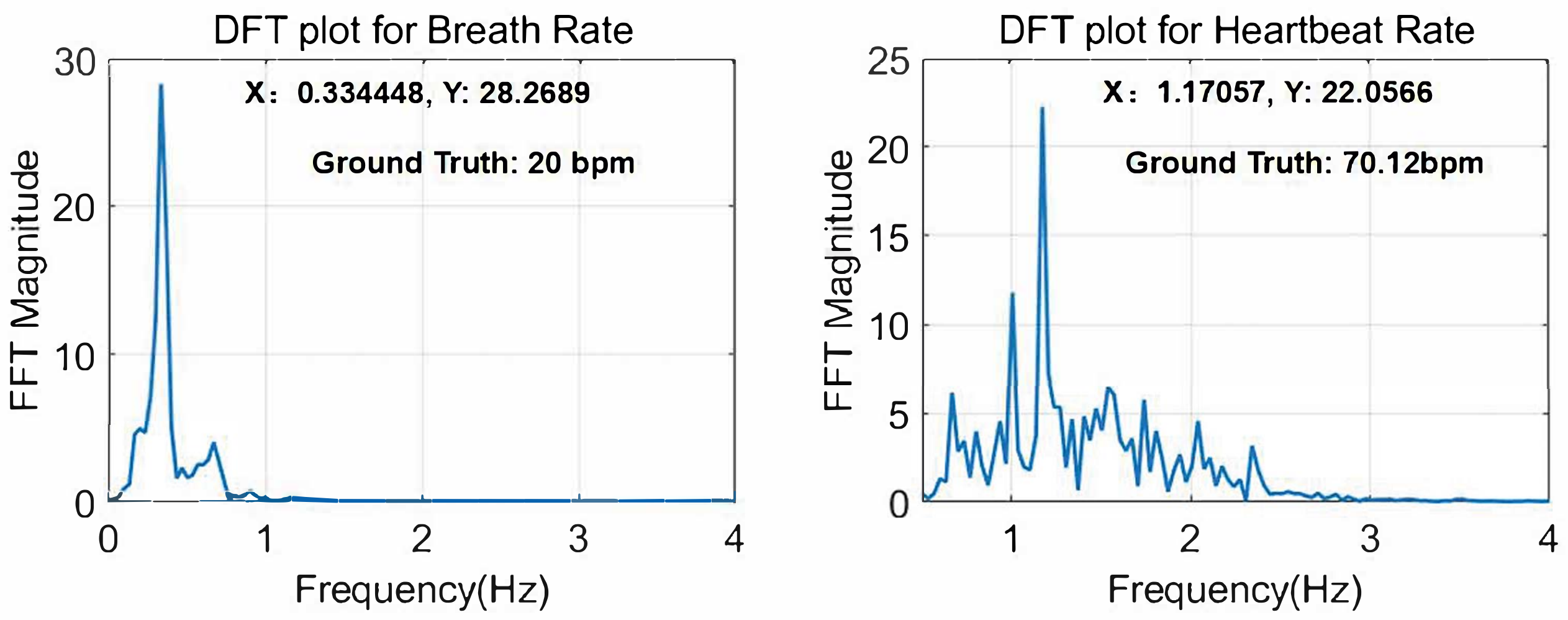}  
    \caption{BR and HR estimation by applying frequency analysis on the phase of mmWave signals}
    \vspace{-2mm}
    \label{fig:fig_fft}
\end{figure}

\begin{center}
\begin{table}[t]%
\caption{Chirp Parameters used in this work} 
\vspace{-1mm}
\centering
\begin{tabular}{|c|c|}
\hline
Start Frequency,$f_{c}$(GHz)  & 77\\\hline
Frequency Slope, S (MHz/$\mu$s)  & 29.982\\\hline
Idle Time ($\mu$s) & 100 \\\hline
TX Start Time ($\mu$s) &0 \\\hline
ADC Start Time($\mu$s) & 6 \\\hline
ADC Samples & 256  \\\hline
ADC Sample Rate (MHz) & 10 \\\hline
Ramp End Time ($\mu$s) & 60 \\\hline
Number of Subframe Per Frame & 2/4 \\\hline
Number of chirp Per Subframe & 1\\\hline
Slow-time Sampling Frequency, $f_{s}$=1/$T_{s}$ (Hz) & 20 \\\hline
Subframe Periodicity (ms)& 12.5\\\hline
Frame Periodicity (ms)& 50 \\\hline
\end{tabular}
\vspace{-1mm}
\label{tab1}
    \label{tab:my_label}
\end{table}
\end{center}
%\vspace{-5mm}
Our system first works in TDM-MIMO mode to detect targeting subjects, leveraging the sparse asymptotic minimum variance (SAMV) algorithm ~\cite{Abeida2013iterative}, which is a super-resolution angle estimation algorithm.
Upon detecting one or more target subjects, our system steers the mmWave beam towards the directions of the detected targets by applying analog beamforming at the Tx side. Our system then utilizes digital beamforming, phased calibration, and frequency analysis techniques as described below to estimate vital signs of the target subjects.  
% Once it detects the targets, based on the estimated angles, the weights of each subarray will be computed and applied to form beams illuminating the targets.
% In order to avoid the interference between different targets,  

\vspace{-2mm}
\subsection{Multiple Channels Combination via Digital Beamforming}
To further enhance the directivity of the radar, we leverage digital beamforming (DBF)~\cite{Barton1980antenna} at the receiver side to make the energy of the echoed mmWave signals focus on the target subject. With the designed TDM phased-MIMO radar, our system fully leverages all TX-RX pairs (i.e., $N\times{M}$) to realize digital beamforming to improve the signal-to-noise ratio. Note that in our radar design, different sets of TX-RX pairs work in a TDM fashion, with only the TX-RX pairs in a subarray activated at one time. Specifically, we compute the beamforming weights at angle $\theta$ as: 
\begin{align}
    w(n,m) = e^{j2\pi(d_m-d_n)\frac{\sin{\theta}}{\lambda}}
\end{align}
where 
$d_n = (n-1){d_t}$ and $d_m = (m-1){d_r}$. With the beamforming weights, we can combine mmWave signals of $k$-th ADC sample in the $l$-th chirp from all Tx-Rx pairs as
\begin{align}
    z[k,l] &= \sum_{n}^{N}\sum_{m}^{M} w(n,m)y[n,m,k,l] \nonumber\\
    &= NMA_r e^{j2\pi[f_bkT_f + \frac{2f_c}{c}R(kT_f + lT_s)]}.
\end{align}
Our system then leverages the combined signal $z$ for BR and HR estimation.  

% The use of digital beamforming can efficiently increase the energy of mmWave signal towards the target's direction.  

% Need to mention that, different subarray transmits in different time slots thus one can separate the signals from different subarrays by time. Then based on the location of illuminated target, the signals from different channels are combine by digital beamforming (DBF), where the phase shift due to the spacing of TX and RX (see \eqref{beat_sig}) can be compensated. 

\vspace{-2mm}
\subsection{Range Detection and Background Noise Cancellation} 
Upon applying range DFT upon the combined signal, our system detects the radial distance between the target and the radar~\cite{Patole2017automotive}. In practical environments, the static objects in the environment may interfere with the detection of the target. To address this problem, we first profile the environmental reflections (i.e., outputs of range DFT without subjects), and then subtract the noises to cancel the impacts of the environment. Such a noise cancellation mechanism renders reliable target detection, even the target is far from the radar device. After the noise cancellation, we select the range bin with the highest range DFT output to represent the distance between the target and the radar. Our system then computes the phase of the range DFT outputs for vital sign estimation, which reflects the distance variations caused by human chest displacements.
     
% After the range DFT and background noise removal, the remaining highest peak indicates the distance of the desired human target and the change of phase in the peaks between different time slots can be used to measure the vital sign of the detected human target.

\vspace{-2mm}
\subsection{Constellation Correction with Least-Square}  
Due to the strong coupling effects and interference in the measurement environment, the range DFT outputs usually contain DC offsets, which distort the phase information at the target's range bin. It is thus necessary to remove the dc offset to derive reliable vital sign information. Given a Tx-Rx pair, $i$, the phase at a selected range bin $h$ can be formulated as
\begin{equation}
\setlength{\abovedisplayskip}{2pt}
\setlength{\belowdisplayskip}{2pt}
\begin{aligned}
\phi(t) = arctan[\frac{Im(r_{i,h}(t)+DC_{im})}{Re(r_{i,h}(t)+DC_{re})}],
\end{aligned}
\label{eq:arctan_phase}
\end{equation}
where $DC_{im}$ and $DC_{re}$ denote the imaginary and real parts of the complex DC offset, respectively. $t$ denotes the time index, and $r_{i,h}(t)$ shows the range DFT output at the range bin $h$. The formulation indicates that the DC offset will shift the origin to $(DC_{re}, DC_{im})$. To compensate such a shift, we leverage least-squares algorithm to estimate and cancel $(DC_{re}, DC_{im})$~\cite{alizadeh2019remote}. 
% and subtract it from the each complex measurement.

% \begin{figure}
%     \centering
%     \includegraphics[width=1.7in]{images/DC_offset}
%     \caption{Constellation correction using least square method}
%     \label{fig:dc_offset}
% \end{figure}

\begin{figure}[t]
    \centering
    \includegraphics[width=0.44\textwidth]{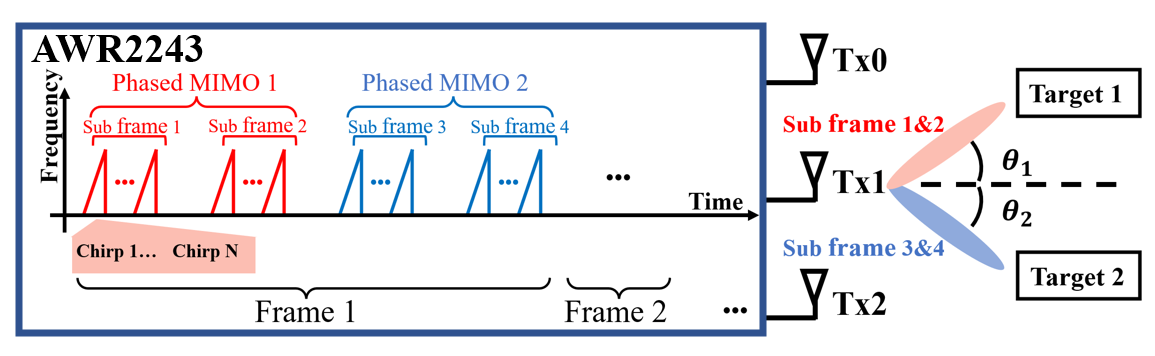}
    \vspace{-1mm}
    \caption{Subframe implementation on AWR2243 (N=1)}
    \vspace{-2mm}
    \label{fig_subframe}
\end{figure}

\vspace{-2mm}
\subsection{Phase Calculation with DACM}
Existing work~\cite{Groote1997chest} found that the human chest displacement can exceed the wavelength of mmWave signals (i.e., $<4$mm for 77GHz). Therefore, the phase of the range DFT can be over the range $[-\pi,\pi]$, which can lead to false detections of vital signs. To tackle this issue, we utilize differential and cross-multiply algorithm (DACM) to calculate phase. 

Instead of directly applying arctangent demodulation, DACM converts complex range DFT outputs of $l-th$ chirp into phases leveraging the derivative of arctangent function
\begin{align}
    \phi(l) &= \phi(l-1) + \Delta\phi(l), \quad l = 2,3,\dots, N_{fra},
\end{align}
where $N_{fra}$ is the number of frames and
% \small{
\begin{equation*}
    \Delta\phi(l) = \frac{Y_R[l]\{Y_I[l]-Y_I[l-1]\}-\{Y_R[l]-Y_R[l-1] \}Y_I[l]}{Y_R[l]^2 + Y_I[l]^2}.
\end{equation*}
The DACM algorithm mainly corrects the phase distortions caused by breathing. In contrast, small-scale heartbeat motions are less likely to exceed the range of phase.  
 
% To estimate such a minor frequency, other post-processing methods were used and explained below to facilitate the heart rate estimation.
% \begin{figure}
%     \centering
%     \includegraphics[width=1.7in]{images/Unwarpped_phase.eps}
%     \caption{Using the interpolation of three previous phases to replace phase which exceed threshold }
%     \label{fig:phase_unwarp}
% \end{figure}

\vspace{-2mm}
\subsection{Phase Drift Calibration based on Phase Difference}
Signal phase drifts in transmission are mainly caused by the impacts of temperature and humidity variations on the hardware, which make the range of phase fluctuations exceed the normal ranges of human breathing and heartbeat. 
% exceed more than one order of magnitude of phase synchronization. 
The phase drifts cannot be removed leveraging DACM, since these drifts can be close to but not exceeding the unwrapping threshold of $\pm\pi$. Furthermore, the harmonics of breathing~\cite{mabrouk2016human} (i.e., multiple of breathing frequency range $0.2-0.33$Hz) can also distort the phase patterns at the heartbeat frequency range (e.g., $0.8-2.0$Hz). It is thus necessary to remove the impacts of such harmonic for reliable HR estimation.    
Particularly, we realize phase drift calibration ~\cite{ahmad2018vital} by computing  the phase difference $\Delta\phi(l)=\phi(l)-\phi(l-1)$ for each $\phi(l)$. If the absolute value of the phase difference exceeds a certain threshold, $\phi(l)$ will be replaced by a new value computed by the Lagrange interpolation using previous three phases $\phi(l-3),\phi(l-2),\phi(l-1)$. 
% if the phase is not in the range of $[-\pi,\pi]$.   

\vspace{-2mm}
\subsection{BR and HR Estimation}
We apply frequency analysis upon the calibrated phase within a sliding window to estimate BR and HR. Since the periods of human breathing and heartbeat are close to each other, we need to separate the BR and HR for reliable estimations. Particularly, we apply a $4$-th order Butterworth bandpass filter with a cut-off frequency range of $0.8-2.0$Hz to extract heartbeats, which removes the impacts of human breathing and its harmonics. Similarly, we use a bandpass filter of $0.1-0.5$Hz to extract human breaths. Then, our system applies DFT on the extracted breathing signals, with the highest peak of the DFT magnitude as the detected BR. To extract heartbeat, which involves subtler displacement, we first calculate phase difference: $\Delta\phi(l)=\phi(l)-\phi(l-1)$, which reveals minor phase changes~\cite{Wang2020RemoteMO}. Our system then applies DFT upon the phase differences to calculate the HR. Examples of BR and HR estimation results are shown in Fig. \ref{fig:fig_fft}.

% An example of breath rate and heartbeat rate estimation applying FFT is shown in Fig. \ref{fig:fig_fft}.

\begin{figure}
    \centering
    \includegraphics[width = 2.4in]{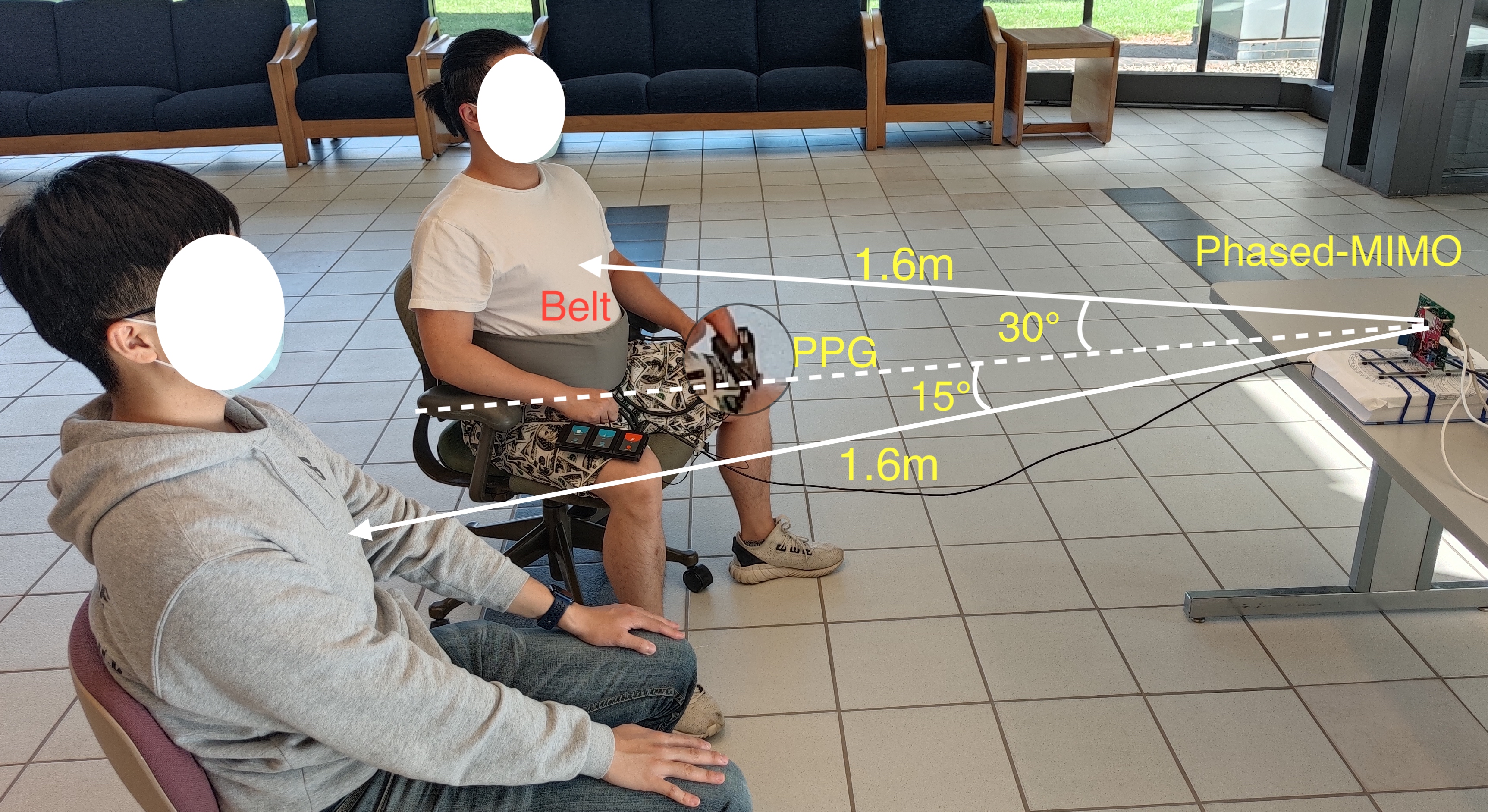} 
    \caption{Setup for multi-target vital sign monitoring.}
     \vspace{-2mm}
    \label{multi setup}
\end{figure}

\vspace{-2mm}
\section{Performance evaluation}
\subsection{TDM-Phased-MIMO Implementation}
As a proof of concept, we implement the proposed TDM phased-MIMO radar on an off-the-shelf Texas Instrument (TI) AWR $2243$ mmWave device~\cite{AWR2243,TIMIMO}, which transmits and receives FMCW waveforms within $76 GHz\sim81 GHz$ frequency range. The mmWave device consists of three TXs with the spacing of $\lambda$ and four RXs with the spacing of $\lambda/2$, respectively. 
% \textcolor{orange}{The RF signal is illuminated towards the target through TXs.} 
% Due to the cardiac activity, the distance between the target and the radar sensor is influenced by the chest displacement. Thus, the RF signal is modulated by the chest motion and then reflected towards RXs. 
An evaluation board TI DCA $1000$ ~\cite{DCA1000}is adopted in the streaming mode to acquire raw baseband I/Q signals down-converted from received signals. It is worth mentioning that when leveraging TXs to realize the analog beamforming, there will be grating lobes pointing to other directions  because of the large spacing between TXs, $\lambda>\lambda/2$, which may introduce interference on the multi-target scenario. Nevertheless, RXs are spaced by $\lambda/2$ which means there is no grating lobes at the receiver side. By leveraging the DBF at receiver side, we can alleviate the grating lobe problem since the energy is focused to a certain direction. In multiple-subject scenarios, our TDM phased-MIMO radar can changes the beam direction towards two different subjects within one frame periodicity of 50 ms. 
% At different measurement times, RF signals radiate along the same direction with different references, TX0 and TX2, for beamforming, respectively.
% To measure multiple targets, the beam direction of transmitted RF signal varies periodically with one frame periodicity of 50 ms, based on FMCW method. 
Each frame is equally divided into four sub frames~\cite{TIprogramming} for multi-target detection, which block diagram is shown in Fig.\ref{fig_subframe}. 
%It can be seen from Fig.\ref{fig_subframe} that 
Particularly, the direction 1 can be illuminated using subframe 1 with TX0 as the reference and subframe 2 with TX2 as the reference, while the direction 2 can be illuminated using subframe 3 with TX0 as the reference and subframe 4 with TX2 as the reference.

\begin{figure}[!t]
\centering
\includegraphics[width= 3in]{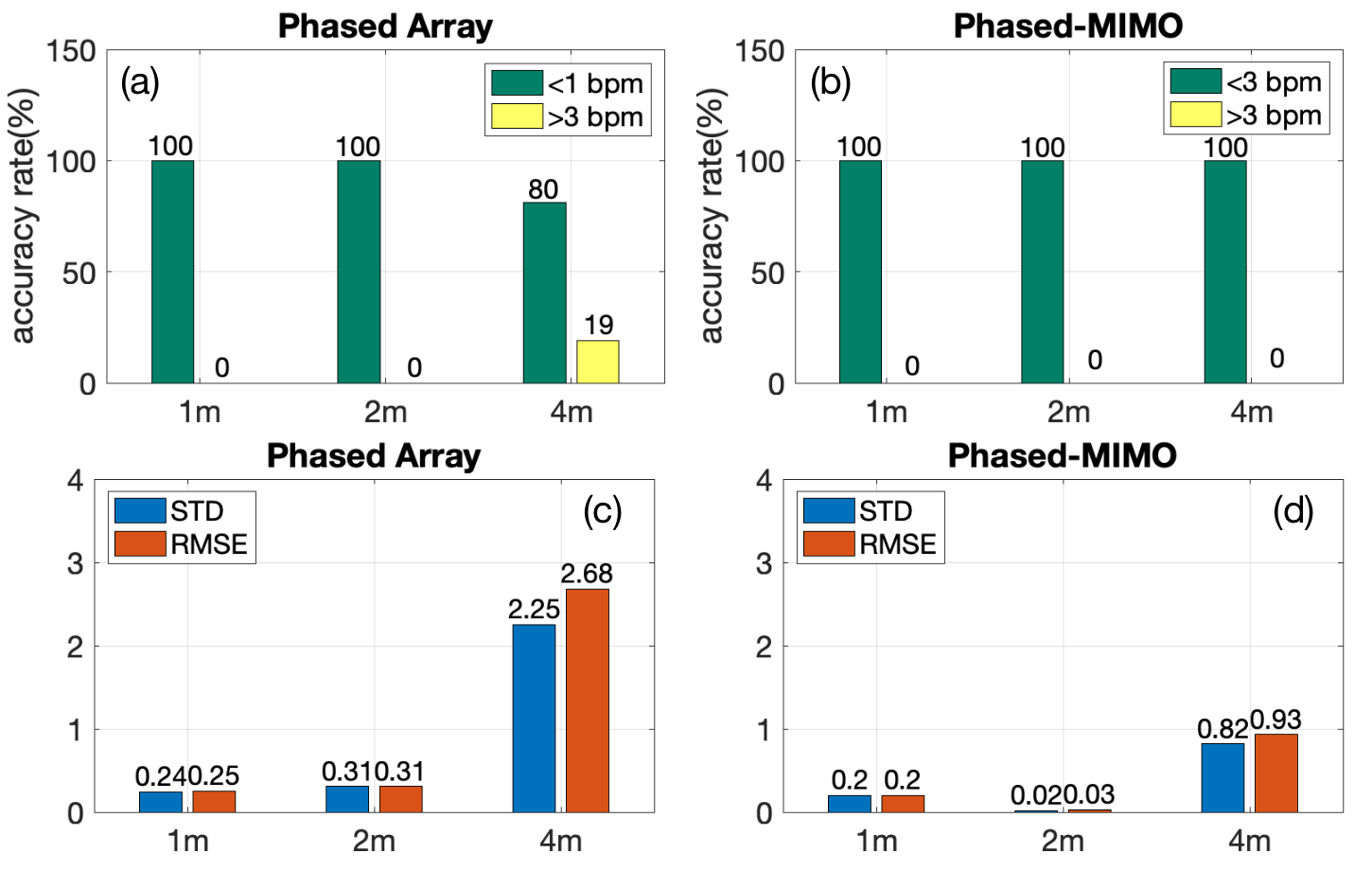}
% \subfloat[]{\includegraphics[width=1.7in]{images/phased_array_br.png}%
% \label{pa_br_single}}
% \hfil
% \subfloat[]{\includegraphics[width=1.7in]{images/phased_mimo_br.png}%
% \label{pm_br_single}}
% \newline
% \subfloat[]{\includegraphics[width=1.7in]{images/phased_array_br_err.png}%
% \label{pa_br_err_single}}
% \hfil
% \subfloat[]{\includegraphics[width=1.7in]{images/phased_mimo_br_err.png}%
\label{pm_br_err_single}
\vspace{-1mm}
\caption{(a),(b) accuracy on single-target BR estimation; (c)(d) corresponding errors.
% X-axis: distance between targets and radar.
% when single target is at a direction of $20.1\degree$ with a distance of $1$,$2$ and $4$ meters, respectively. 
X-axis: subject-to-radar distance.
% and the y-axis is the accuracy. In the experiments, human targets are required to breathe normally. respectively. The y-axis indicates the error level.
% \textcolor{red}{BR result}, like the "Enhancement..." paper, put all the captions of subfigures here. what is the x-axis, experiment setup: single target, 1m/2m/4m, 20.1\degree, breath normally, multiple person involved.
}
\vspace{-5mm}
\label{br_single}
\end{figure}

\vspace{-2mm}
\subsection{Experimental Validation and Error Analysis}
We evaluate the performance of our vital sign monitoring system under single-subject and two-subject scenarios. For both scenarios, we conduct experiments to study the impacts of various factors, including the distance between the radar and the subject and the separation angle between two subjects. Each experiment lasts 2 minutes. We use breathing and heartbeat signals collected with a Neulog NUL236 respiration belt and a Neulog NUL208 Heart Rate sensor as the ground truth. A 60-second sliding window, with a step size of 1 second, is applied upon the breathing and heartbeat signals to obtain the ground-truth BR and HR. We compare the HR and BR estimated with our system based on the same sliding window with the ground truth for error analysis. To quantify the vital sign estimation performance, we use statistical metrics including standard derivation (STD), root-mean-square error (RMSE), and estimation accuracy, which are also exploited in prior work~\cite{alizadeh2019remote}. Specifically, STD indicates the consistency of the estimations, and a lower STD means higher consistency and better performance. RMSE measures the average errors between the estimations of our system and the ground truth. Besides these two statistical metrics, we use estimation accuracy for evaluation, which is defined as the percentage of the estimation with $<3$ bpm errors. 
 
\begin{figure}[!t]
\centering
\includegraphics[width=3in]{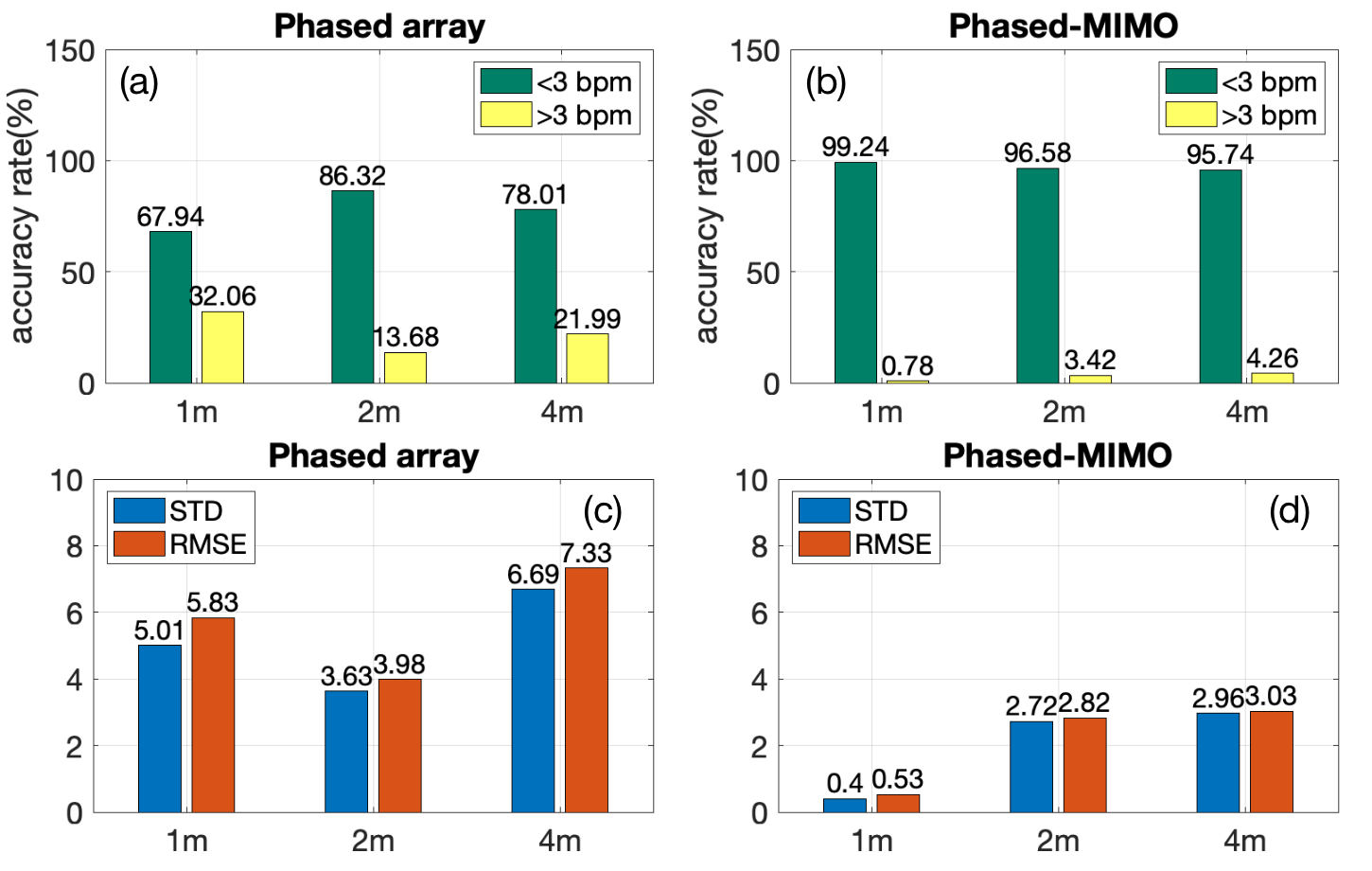}
% \subfloat[Case I]{\includegraphics[width=1.5in]{images/phased_array_hr.png}%
% \label{pa_hr_single}}
% \hfil
% \subfloat[Case II]{\includegraphics[width=1.5in]{images/phased_mimo_hr.png}%
% \label{pm_hr_single}}
% \newline
% \subfloat[Case I]{\includegraphics[width=1.5in]{images/phased_array_hr_err.png}%
% \label{pa_hr_err_single}}
% \hfil
% \subfloat[Case II]{\includegraphics[width=1.5in]{images/phased_mimo_hr_err.png}%
% \label{pm_hr_err_single}}
\vspace{-1mm}
\caption{(a),(b) accuracy on single-target HR estimation; (c)(d) corresponding errors.
% X-axis: distance between target and radar.
% when single target is at a direction of $20.1\degree$ with a distance of $1$,$2$ and $4$ meters, respectively. 
X-axis: subject-to-radar distance. 
% \ref{pa_hr_single} and \ref{pm_hr_single} show the accuracy of phased array and phased-MIMO on HR estimation when single target is at a direction of $20.1\degree$ with a distance of $1$,$2$ and $4$ meters, respectively. The x-axis denotes the distance of target to radar sensors and the y-axis is the accuracy. In the experiments, human targets are required to breathe normally.\ref{pa_hr_err_single} and \ref{pm_hr_err_single} are the corresponding error of phased array and phased-MIMO, respectively. The y-axis indicates the error level.
% \textcolor{red}{HR result}, like the "Enhancement..." paper, put all the captions of subfigures here. what is the x-axis, experiment setup: single target, 1m/2m/4m, 20.1\degree, breath normally, multiple person involved.{phased array perform bad at 1m due to bad channel. only channel 1 is used.}
}
\vspace{-5mm}
\label{hr_single}
\end{figure}

%%\subsection{Implement Phased-MIMO on 2243} %%Merge into the previous subsection 
\vspace{-2mm}
\subsection{Single-target Vital Sign Estimation} 

In the scenario of single-target vital sign monitoring, the subject is requested to sit in front of the radar, which is placed 1m, 2m and 4m away from the subject with a direction of $20.1\degree$. The estimation accuracy and statistic results of single-target BR are shown in Fig. \ref{br_single}. For the single-target BR estimation on 1m and 2m, the estimation accuracy for phased-array can reach 97\%. However, under a far radar-to-subject distance of 4m, the accuracy drops to $80\%$ for predictions of $<3 bpm$ errors.
% But for the setup of 4m between the subject and radar, the accuracy of single-target breath rate estimation can only achieve an accuracy of 57.9\% with the error lower than 1 bpm. 
In contrast, the phased-MIMO not only maintains high accuracy on BR estimation on 1m (99\%) and 2m (100\%), but achieve much higher accuracy at $4m$, with $100\%$ for predictions of $<3 bpm$ errors. For the statistic results, the STD and RMSE of both phased-array and phased-MIMO will experience a severe increasing when the distance between subjects and radar changes to 4m, which can attribute to the instability of long distance data collection from radar. Compared to phased-array, phased-MIMO has much lower STD and RMSE on all three different distances, demonstrating the superior performance on BR estimation. 

In Fig.~\ref{hr_single}, we compare the performance of phased array and proposed phased-MIMO HR estimations. 
% We can also find the effectiveness and robustness using phased-MIMO according to the results of single-target heartbeat rate estimation using phased-array and phased-MIMO in Fig. \ref{hr_single}. 
For the phased-array, the accuracies are only $67.9\%$, $86.3\%$, and $78.0\%$, on 1m, 2m, and 4m. 
In contrast, the system with phased-MIMO has much better HR estimation performance, with over $95\%$ accuracy on three different distances. We note that phased array has the best performance on the setup of 2m, with the STD of 3.63 and RMSE of 3.98. The possible reason of this result is that the Radar Cross-Section is low due to the shorter distance between the subject and radar, and data collection on radar's side is not stable if the distance is longer. For phased-MIMO, even in longer distance, it still have lower STD and RMSE, which proves the stability and robustness using phase-MIMO to make single-target heartbeat estimation.

\begin{figure}[!t]
\centering
\includegraphics[width=3.12in]{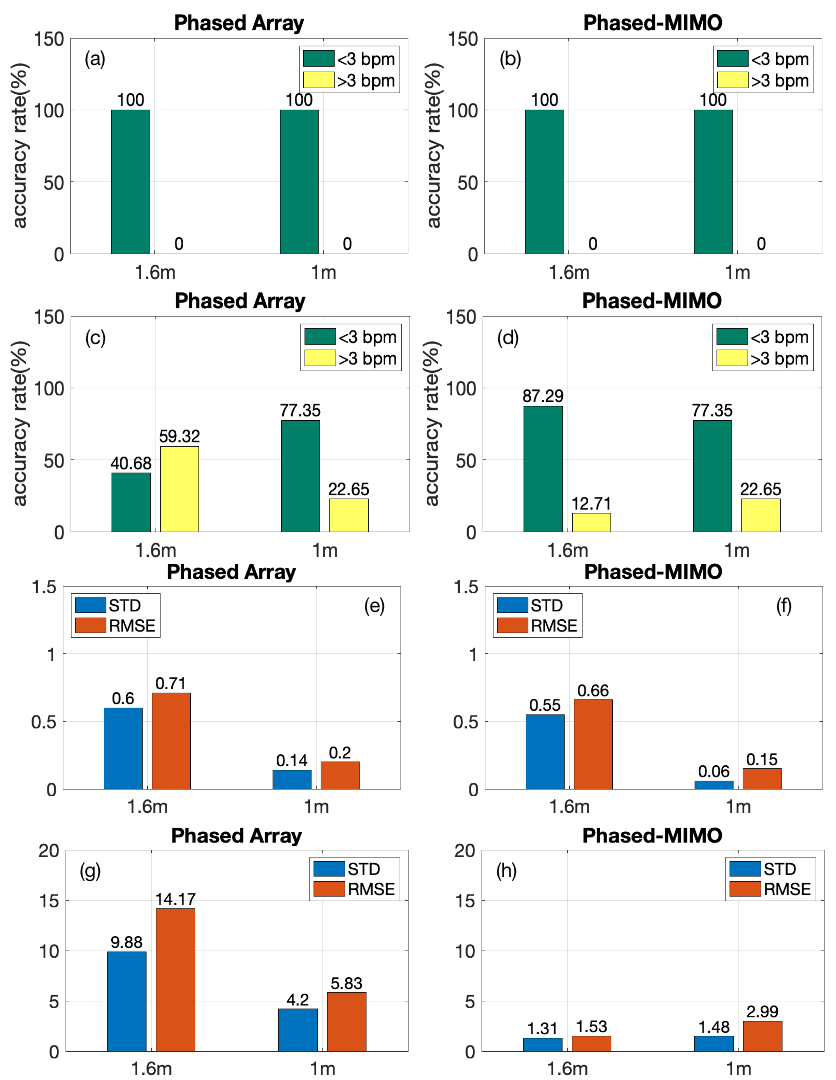}
% \subfloat[Case I]{\includegraphics[width=1.5in]{images/fixed_separation/phased_array_br.png}%
% \label{pa_br_distance}}
% \hfil
% \subfloat[Case II]{\includegraphics[width=1.5in]{images/fixed_separation/phased_mimo_br.png}%
% \label{pm_br_distance}}
% \newline
% \subfloat[Case I]{\includegraphics[width=1.5in]{images/fixed_separation/phased_array_br_err.png}%
% \label{pa_br_distance_err}}
% \hfil
% \subfloat[Case II]{\includegraphics[width=1.5in]{images/fixed_separation/phased_mimo_br_err.png}%
% \label{pm_br_distance_err}} 
\vspace{-1mm}
\caption{(a)(b) accuracy on multi-target BR estimation; (c)(d) accuracy on multi-target HR estimation; (e)(f) errors for multi-target BR estimation; (g)(h) errors for multi-target HR estimation.
% X-axis: distance between targets and radar.
% when single target is at a direction of $20.1\degree$ with a distance of $1$,$2$ and $4$ meters, respectively. 
X-axis: subject-to-radar distance. The angle separation between subjects is $40\degree$.
% \ref{pa_br_distance} and \ref{pm_br_distance} show the accuracy of phased array and phased-MIMO on BR estimation when two targets are sitting in front of radar with a fixed angle separation of $60\degree$, i.e., $30\degree$ and $-30\degree$.
% Two targets are in the same range bin with a distance of $1$ and $1.6$ meters, respectively. The x-axis denotes the distance of targets to radar sensors and the y-axis is the accuracy. In the experiments, human targets are required to breathe normally and in different rates following two metronomes.\ref{pa_br_distance_err} and \ref{pm_br_distance_err} are the corresponding error of phased array and phased-MIMO, respectively. The y-axis indicates the error level.
% \textcolor{red}{Multi-target BR, fixed angle separation with 60°, (-30 ,30) distance are 1m and 1.6m}, like the "Enhancement..." paper, put all the captions of subfigures here. what is the x-axis, experiment setup: single target, 1m/2m/4m, 20.1\degree, breath normally, multiple person involved.
}
\vspace{-5mm}
\label{fig_sim_3}
\end{figure}

\vspace{-2mm}
\subsection{Multi-target Vital Sign Estimation}
Multi-target vital signs detection is more challenging, especially when the targets are in the same range bin since FMCW signal only tells the range information~\cite{alizadeh2019remote}. For that case, we propose to use analog and digital beamforming to separate different targets in angle domain. By transmitting beams towards each of the target, we can isolate the targets at the same distance; by DBF, we can further focus the energy of received signal towards the desired target and thus reduce the interference from other targets. In the first multi-target detection experiment, two targets are sitting along the direction of $-30 \degree$ and $30 \degree$ with the distance of $1$m away from the radar sensor. In this case, the phase angles provided by phase shifters for TX0, TX1 and TX2 are $0 \degree$, $180 \degree$ and $0 \degree$, respectively. Due to the large spacing between TXs, the grating lobe issue appears along the direction of $-30 \degree$ with the maximum gain of main beam along the direction of $30 \degree$ and vice versa. Fortunately, we can use DBF to address this problem.

The statistic result comparison between phased array and phased-MIMO are shown in Fig. \ref{fig_sim_3} when two targets located at different distances with the same angle separation of $60 \degree$. In Fig. \ref{fig_sim_3}(a) and (b), for BR measurement, both the accuracy rates for phased-MIMO and phased array are 100 $\%$. At the distance of 1 m, the STD and RMSE for phased array are 0.14 and 0.2, respectively, while the STD and RMSE for phased-MIMO are 0.06 and 0.15, respectively. At the distance of 1.6 m, the STD and RMSE for phased array are 0.6 and 0.71, respectively, while the STD and RMSE for phased-MIMO are 0.55 and 0.66, respectively. For HR measurement in Fig. \ref{fig_sim_3}(c) and (d), the accuracy rates between phased array and phased-MIMO are 40.68 $\%$ and 87.29 $\%$, At the distance of 1 m, the STD and RMSE for phased array are 4.2 and 5.83, respectively, while the STD and RMSE for phased-MIMO are 1.48 and 2.99, respectively.At the distance of 1.6 m, the STD and RMSE for phased array are 9.88 and 14.17, respectively, while the STD and RMSE for phased-MIMO are 1.31 and 1.53, respectively.

\begin{figure}[!t]
\centering
\includegraphics[width=3.08in]{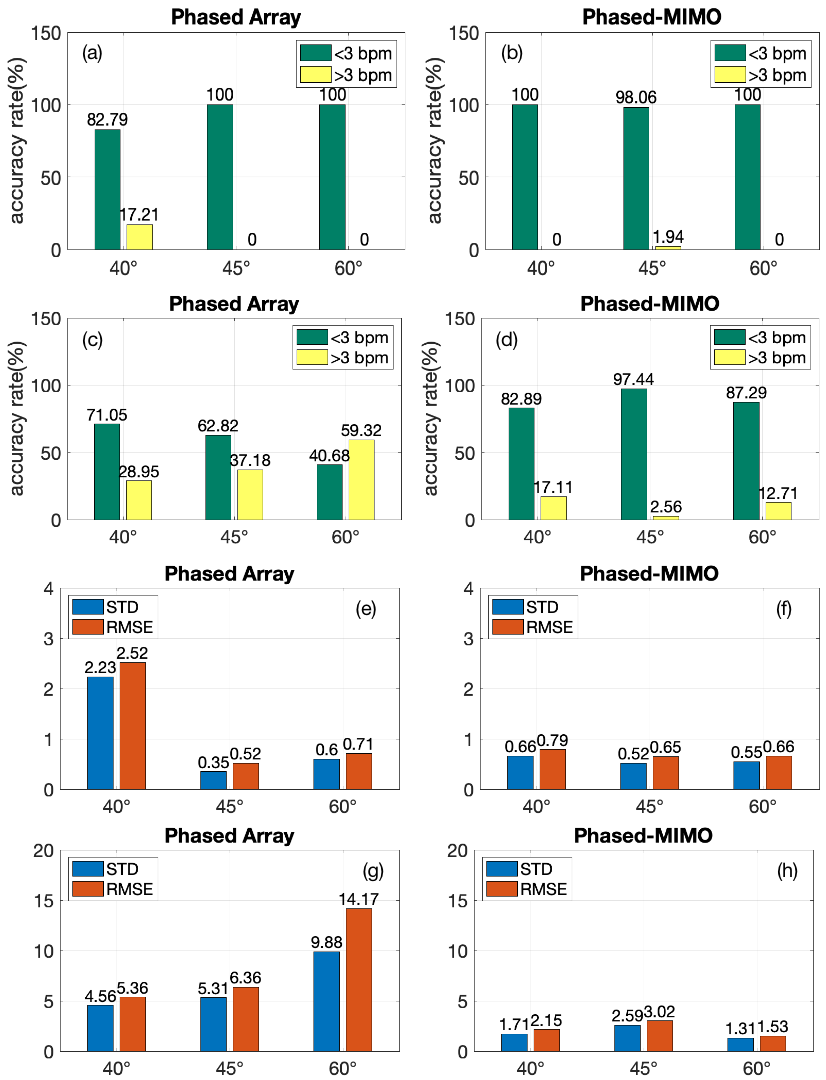}
% \subfloat[Case I]{\includegraphics[width=1.5in]{images/fixed_distance/BR_pa.png}%
% \label{pa_br_angle}}
% \hfil
% \subfloat[Case II]{\includegraphics[width=1.5in]{images/fixed_distance/BR_pm.png}%
% \label{pm_br_angle}}
% \newline
% \subfloat[Case I]{\includegraphics[width=1.5in]{images/fixed_distance/BR_err_pa.png}%
% \label{pa_br_angle_err}}
% \hfil
% \subfloat[Case II]{\includegraphics[width=1.5in]{images/fixed_distance/BR_err_pm.png}%
% \label{pm_br_angle_err}}
\vspace{-1mm}
\caption{(a)(b) accuracy on multi-target BR estimation; (c)(d) accuracy on multi-target HR estimation; (e)(f) errors for multi-target BR estimation; (g)(h) errors for multi-target HR estimation.
X-axis: angle between two targets. The radar-to-target distances are all $1.6m$.
% \ref{pa_br_angle} and \ref{pm_br_angle} show the accuracy of phased array and phased-MIMO on BR estimation when two targets are sitting in front of radar with a fixed distance of $1.6$meters while the first target sitting at $30\degree$ and the other target sitting at $-30\degree$, $-15\degree$, and $-10\degree$, respectively.
% The x-axis denotes the angle separation between two targets and the y-axis is the accuracy. In the experiments, human targets are required to breathe normally and in different rates following two metronomes.\ref{pa_br_angle_err} and \ref{pm_br_angle_err} are the corresponding error of phased array and phased-MIMO, respectively. The y-axis indicates the error level.
% \textcolor{red}{BR result, fixed distance 1.6m, target 1 at 30°, target 2 at -30, -15, -10 degrees.}, like the "Enhancement..." paper, put all the captions of subfigures here. what is the x-axis, experiment setup: single target, 1m/2m/4m, 20.1\degree, breath normally, multiple person involved.
}
\vspace{-4mm}
\label{multi_distance}
\end{figure}

When two targets located at the distance of 1.6 m with different angle separations of $40 \degree$, $45 \degree$ and $60 \degree$, the performance of phased array and phased-MIMO are summarized in Fig. \ref{multi_distance}. In Fig. \ref{multi_distance}(a) and (b), for BR measurement, the accuracy rate for phased array is larger than 82.79 $\%$ while the accuracy rate for phased-MIMO is larger than 98.06 $\%$ when the angle separation is changed from $40 \degree$ to $60 \degree$. As shown in Fig. \ref{multi_distance} (e) and (f), the STD and RMSE results for phased-MIMO are smaller than those for phased array. Similar results can be seen in Fig. \ref{multi_distance}(g) and (h). For HR measurement, the accuracy rates for phased-MIMO are larger than those for phased array under different angle separations. 

% The STD and RMSE results for phased-MIMO is smaller than those for phased array.
% {Yichao: why is the STD and RMSE for phased MIMO is larger than those for phased array with angle separation of 45 degree in Fig. 8 (c) and (d).}

% Describe figures. Compare. How we separate two person's breathe rate( m

\vspace{-1mm}
\section{Conclusion}
We designed a TDM phased-MIMO radar to realize high-precision multi-people vital sign monitoring. The designed radar can successively steer the mmWave beam towards different directions, which enables the separation of vital signals of multiple people at the same range bin and integrates the MIMO technique into our design to boost the SNR.
% Furthermore, for each direction, we integrated the MIMO technique into our design to boost the SNR and obtain fine-grained measurements, through transmitting waveforms with multiple sub-array in a TDM fashion. 
Furthermore, we developed a system to localize multiple target subjects and extract echoed mmWave signals of each individual subject. Our system then computes the phases of the echoed signals and applies a set of calibration and processing techniques for reliable BR and HR estimation. As compared to the phased array, our TDM-phased MIMO can more accurately estimate the BR and the HR of multiple subjects.

\bibliographystyle{IEEEtran}
\bibliography{reference.bib}

\end{document}